\documentclass[11pt]{article}
\pdfoutput=1

\usepackage{amsmath,hyperref}
\usepackage{fullpage}

\newtheorem{theorem}{Theorem}

\begin{document}

\title{The Nelson-Seiberg theorem generalized\\
       with nonpolynomial superpotentials}
\author{Zhengyi Li\textsuperscript{*}, Zheng Sun\textsuperscript{\dag}\\
        \normalsize\textit{College of Physics, Sichuan University, Chengdu 610064, P. R. China}\\
        \normalsize\textit{E-mail:}
        \textsuperscript{*}\texttt{897237485@qq.com,}
        \textsuperscript{\dag}\texttt{sun\_ctp@scu.edu.cn}
       }
\date{}
\maketitle

\begin{abstract}

The Nelson-Seiberg theorem relates R-symmetries to F-term supersymmetry breaking, and provides a guiding rule for new physics model building beyond the Standard Model.  A revision of the theorem gives a necessary and sufficient condition to supersymmetry breaking in models with polynomial superpotentials.  This work revisits the theorem to include models with nonpolynomial superpotentials.  With a generic R-symmetric superpotential, a singularity at the origin of the field space implies both R-symmetry breaking and supersymmetry breaking.  We give a generalized necessary and sufficient condition for supersymmetry breaking which applies to both perturbative and nonperturbative models.

\end{abstract}

\section{Introduction}

Supersymmetry (SUSY)~\cite{Nilles:1983ge, Martin:1997ns, Wess:1992cp, Bailin:1994qt, Terning:2006bq, Dine:2007zp} provides a natural solution to several unsolved problems in the Standard Model (SM), through its extension to the supersymmetric Standard Model (SSM).  In this framework, bosons and fermions appear in pairs related by SUSY\@.  So every particle in SM has a SUSY partner called a sparticle, which has similar properties to its corresponding SM particle.  The mass spectrum of sparticles will be the same as SM particles if SUSY is a good symmetry at low energy.  Since sparticles have not been discovered yet, SUSY must be broken to give them heavy masses escaping the current experimental limit~\cite{Tanabashi:2018oca}.  To avoid the problem of light sparticles in model building, SUSY must be broken in a hidden sector~\cite{Intriligator:2007cp} which introduces new fields beyond SM, and then the SUSY breaking effects are mediated to the observable SSM sector by a messenger sector, giving sparticle mass spectrum and coupling constants which may be examined in future experiments.  There are two types of SUSY breaking models called F-term and D-term SUSY breaking.  D-term SUSY breaking, usually involving Fayet-Iliopoulos terms~\cite{Fayet:1974jb}, has more difficulties to give appropriate sparticle masses and to be consistent with quantum gravity~\cite{Komargodski:2009pc, Dienes:2009td, Komargodski:2010rb}.  So this work focuses on F-term SUSY breaking, assuming vanishing D-terms at the vacuum.

F-term SUSY breaking models, also called Wess-Zumino models~\cite{Wess:1973kz, Wess:1974jb} or O'Raifeartaigh models~\cite{ORaifeartaigh:1975nky}, involve superpotentials which are holomorphic functions of chiral superfields.  In SUSY model buiding, R-symmetries are often utilized because of the relation between R-symmetries and SUSY breaking discovered by Nelson and Seiberg~\cite{Nelson:1993nf}.  Metastable SUSY breaking~\cite{Intriligator:2006dd} also benefits from approximate R-symmetries through an approximate version of the Nelson-Seiberg theorem~\cite{Intriligator:2007py, Abe:2007ax}.  A revised version of the Nelson-Seiberg theorem gives a combined necessary and sufficient condition for SUSY breaking with an assumption of generic polynomial superpotentials~\cite{Kang:2012fn}, while the original theorem applies to any generic superpotentials but gives separate necessary and sufficient conditions.  Although counterexamples with generic coefficients are found \cite{Sun:2019bnd, Amariti:2020lvx}, they have nongeneric R-charge assignments so that do not violate both the original and the revised theorems.  This work extends the previous analysis to cover models with nonpolynomial superpotentials which are often found in dynamical SUSY breaking.  We give a generalized theorem on a necessary and sufficient condition for SUSY breaking in both models with polynomial and nonpolynomial superpotentials.  It provides a guiding rule for low energy effective SUSY model building to study new physics beyond SM\@.

The rest part of this paper is arranged as following.  Section 2 reviews the original Nelson-Seiberg theorem.  Section 3 reviews a revision of the Nelson-Seiberg theorem which gives a necessary and sufficient condition for SUSY breaking with the assumption of polynomial superpotentials.  Section 4 gives a proof for our generalized theorem covering both models with polynomial and nonpolynomial superpotentials.  Section 5 makes the conclusion and final remarks.

\section{The Nelson-Seiberg theorem}

This section reviews the original Nelson-Seiberg theorem and its proof~\cite{Nelson:1993nf}.  The setup is on a Wess-Zumino model~\cite{Wess:1973kz, Wess:1974jb} which involves a superpotential $W(\phi_i)$ as a holomorphic function of chiral superfields $\phi_i, \ i = 1, \dotsc, d$, and a K\"ahler potential $K(\phi_i^*, \phi_j)$ as a real and positive-definite function of $\phi_i$'s and their conjugates $\phi_i^*$'s.  We use Einstein notation to sum up terms with repeated indices throughout this work.  Although a minimal K\"ahler potential $K(\phi_i^*, \phi_j) = \phi_i^* \phi_i$ is often assumed, most of our analysis in this work is valid for generic K\"ahler potentials.  Since the vacuum is determined by the scalar components $z_i$'s of $\phi_i$'s once the auxiliary components $F_i$'s are solved, $W$ and $K$ are also viewed as functions of $z_i$'s and $z_i^*$'s.  A vacuum corresponds to a minimum of the scalar potential $V$, which is defined as
\begin{equation}
    \begin{gathered}
        V = K_{\bar{i} j} (\partial_i W)^* \partial_j W, \\
        \text{where} \quad
        K_{\bar{i} j} K^{\bar{i} j'} = \delta_j^{j'}, \quad
        K^{\bar{i} j} = \partial_{\bar{i}} \partial_j K, \quad
        \partial_i = \frac{\partial}{\partial z_i}, \quad
        \partial_{\bar{i}} = \frac{\partial}{\partial z_i^*}.
    \end{gathered}
\end{equation}
Whether SUSY breaking happens or not can be checked by solving the F-term equations
\begin{equation}
F_i = \partial_i W = 0.
\end{equation}
A solution to $\partial_i W = 0$ gives a global minimum of $V$ which preserves SUSY\@.  Nonexistence of such a solution can be taken as the criteria for SUSY breaking, although the existence of a SUSY breaking vacua needs to be confirmed by minimizing the scalar potential $V$.  For now we just assume a global minimum exists in models under discussion.  Following the work of Nelson and Seiberg, we are to discuss the existence of a solution to $\partial_i W = 0$, given $W$ with generic terms and coefficients respecting symmetries in each of the following cases.

\begin{itemize}
    \item When there is no R-symmetry, a solution to $\partial_i W = 0$ generically exists, because there are equal numbers of equations and variables.  Introducing a non-R symmetry does not change the situation, because it reduces both equations and variables by a same number.
    \item When there is an R-symmetry, $W$ must have R-charge $2$ in order to make the Lagrangian R-invariant.  So there is at least one field with a nonzero R-charge.  One can choose such a field $z_d$.  With a field redefinition,  $W$ is written as
          \begin{equation} \label{eq:2-01}
              \begin{gathered}
                  W = x f(y_1, \dotsc, y_{d - 1}),\\
                  \text{where} \quad
                  x = z_d ^ {2 / r_d}, \quad
                  y_i = z_i / z_d ^ {r_i / r_d}, \quad
                  i = 1, \dotsc, d - 1,
              \end{gathered}
          \end{equation}
          and $r_i$'s are R-charges of $z_i$'s.  The redefinition makes $x$ to have R-charge $2$ and $y_i$'s to have R-charge $0$.  We consider the following two types of vacua:
          \begin{itemize}
              \item For a vacuum with $x \ne 0$, equations $\partial_i W = 0$ become
                    \begin{equation}
                        \begin{aligned}
                                           f &= 0,\\
                            \partial_{y_i} f &= 0, \quad
                            i = 1, \dotsc, d - 1.
                        \end{aligned}
                    \end{equation}
                    There are $d - 1$ variables to solve $d$ equations, which are overdetermined.  A generic function $f$ does not allow such a solution to exist.  So if such a vacuum with $x \ne 0$ does exist, it generically breaks SUSY\@.
              \item For a vacuum with $x = 0$, equations $\partial_i W = 0$ become
                    \begin{equation}
                        \begin{aligned}
                                             f &= 0,\\
                            x \partial_{y_i} f &= 0, \quad
                            i = 1, \dotsc, d - 1.
                        \end{aligned}
                    \end{equation}
                    The single equation $f = 0$ can always be solved for a generic function $f$, and other equations are all satisfied at $x = 0$.  But the field redefinition in~\eqref{eq:2-01} is usually singular at $x = 0$ except for some special choices of R-charge assignments.  So the existence of a vacuum with $x = 0$ is unclear.
          \end{itemize}
    \item Notice that a vacuum with $x \ne 0$ spontaneously breaks the R-symmetry, while a vacuum with spontaneous R-symmetry breaking means that there is at least one field $z_d \ne 0$ with $r_d \ne 0$, which can be used to make the redefinition in~\eqref{eq:2-01} with $x \ne 0$.
\end{itemize}

In summary, we have proved the original Nelson-Seiberg theorem:

\begin{theorem}
(The Nelson-Seiberg theorem)
In a Wess-Zumino model with a generic superpotential, assuming the existence of a vacuum at the global minimum of the scalar potential, an R-symmetry is a necessary condition, and a spontaneously broken R-symmetry is a sufficient condition for SUSY breaking at the vacuum.
\end{theorem}

\section{The revised theorem with polynomial superpotentials}

This section reviews the revised version of the Nelson-Seiberg theorem and its proof~\cite{Kang:2012fn}.  To avoid singularities in the field space and other complications from the field redefinition in~\eqref{eq:2-01}, we consider the superpotential without doing any field redefinition in the following proof.

\begin{itemize}
    \item When there is no R-symmetry, SUSY is generically unbroken according to the original Nelson-Seiberg theorem.
    \item When there is an R-symmetry, fields can be classified to three types according to their R-charges:
          \begin{equation} \label{eq:3-01}
              \begin{aligned}
                  r(X_i) &= 2, \quad
                  i = 1, \dotsc , N_X,\\
                  r(Y_j) &= 0, \quad
                  j = 1, \dotsc , N_Y,\\
                  r(A_k) &\ne 2, 0, \quad
                  k = 1, \dotsc , N_A.
              \end{aligned}
          \end{equation}
          Assuming the superpotential $W$ has a polynomial form, we can write down the generic form of $W$ by including all monomial combinations of fields with R-charge $2$:
          \begin{equation} \label{eq:3-02}
              \begin{aligned}
                    W =& X_i f_i(Y_j) + W_1,\\
                  W_1 =& \mu_{i j k} X_i X_j A_k
                         + \nu_{i j k} X_i A_j A_k
                         + \xi_{i j k} Y_i A_j A_k
                         + \kappa_{i j} A_i A_j
                         + \lambda_{i j k} A_i A_j A_k +\\
                       & + (\text{nonrenormalizable terms}).
              \end{aligned}
          \end{equation}
          Note that not all $A_i$'s can appear in every terms of $W_1$.  Only those field combinations with R-charge $2$ contribute to $W_1$ with nonzero coefficients.  Each term of $W_1$ contains at least two $X_i$'s or $A_i$'s.  This feature is also possessed by nonrenormalizable terms of $W_1$.  We consider the following cases:
          \begin{itemize}
              \item In the case of $N_X \le N_Y$, setting $X_i = A_i = 0$ makes all first derivatives of $W_1$ equal zero, then solving $f_i(Y_j) = 0$ gives a SUSY vacuum.  Such a solution generically exists because the number of equations, which equals $N_X$, is less than or equal to $N_Y$, the number of variables.
              \item In the case of $N_X > N_Y$, we consider the following two types of vacua:
                    \begin{itemize}
                        \item For a vacuum with $X_i = A_i = 0$, all first derivatives of $W_1$ are set to zero.  But generically there is no solution to $f_i(Y_j) = 0$ because the number of equations is greater than  the number of variables.  SUSY is generically broken if such a vacuum does exist.
                        \item For a vacuum with some $X_i \ne 0$ or $A_i \ne 0$, which carries a nonzero R-charge, the R-symmetry is spontaneously broken by this field.  Then SUSY is generically broken according to the original Nelson-Seiberg theorem.
                    \end{itemize}
              \item If there are more than one consistent R-charge assignments, one should explore all possibilities of R-charge assignments to see whether $N_X \le N_Y$ can be satisfied with one assignment.  SUSY is broken only if $N_X > N_Y$ is satisfied for all possible consistent R-charge assignments.
          \end{itemize}
\end{itemize}

These exhaust all cases with and without R-symmetries.  In summary, we have proved a necessary and sufficient condition for SUSY breaking:

\begin{theorem}
(The Nelson-Seiberg theorem revised)
In a Wess-Zumino model with a generic polynomial superpotential, assuming the existence of a vacuum at the global minimum of the scalar potential, SUSY is spontaneously broken at the vacuum if and only if the superpotential has an R-symmetry and the number of R-charge $2$ fields is greater than the number of R-charge $0$ fields for any possible consistent R-charge assignment.
\end{theorem}

The extra freedom to assign different R-charges can be viewed as a non-R $U(1)$ symmetrie in addition to the R-symmetry~\cite{Komargodski:2009jf}.  $A_k$'s, the fields with R-charges other than $2$ and $0$, do not appear in the SUSY breaking condition of the revised theorem, but are needed for spontaneous R-symmetry breaking to generate gaugino masses~\cite{Shih:2007av, Carpenter:2008wi, Sun:2008va, Curtin:2012yu, Liu:2014ida}.  In addition, according to the above proof procedure, A SUSY vacuum from a model with an R-symmetry and $N_X \le N_Y$ also preserves the R-symmetry and gives a zero expectation value to $W$~\cite{Sun:2011fq}.  Such vacua play important roles in string phenomenology~\cite{Dine:2004is, Dine:2005yq, Dine:2005gz}.

\section{Generalization to include nonpolynomial superpotentials}

The generic form of the superpotential~\eqref{eq:3-02} is an essential step of the previous proof, which comes from the assumption of an R-symmetry and a polynomial $W$.  So a superpotential beyond the polynomial expansion may invalidates the proof of the revised theorem.  But the proof of the original Nelson-Seiberg theorem does not rely on the polynomial form of $W$.  Models in the scope of the original Nelson-Seiberg theorem but out of the scope of the revised theorem often appear in dynamical SUSY breaking.  To achieve a more general theorem to cover these models, we need to analyze $W$ as an arbitrary generic function of fields.

\begin{itemize}
    \item When there is no R-symmetry, SUSY is generically unbroken according to the original Nelson-Seiberg theorem.
    \item When there is an R-symmetry, we suppose fields are properly defined so that the origin of the field space preserves the R-symmetry.  Thus every field transforms by a complex phase angle under the R-symmetry and can be assigned an R-charge.  Fields can be classified to $X_i$'s, $Y_i$'s or $A_i$'s according to their R-charges, just like what has been done in~\eqref{eq:3-01}.  There are $N_Y$ degrees of freedom to choose the origin because any expectation values of $Y_i$'s are invariant under the R-symmetry.  The superpotential $W$ is supposedly a generic holomorphic function of fields.  We consider the following two cases:
          \begin{itemize}
          \item If $W$ is smooth at the origin, it has a Taylor series expansion with a nonzero radius of convergence.  The expansion only needs to be done in variables $X_i$'s and $A_i$'s, and all constant coefficients can be replaced with arbitrary functions of $Y_i$'s.  The generic expansion from the origin $X_i = A_i = 0$ is
                \begin{equation} \label{eq:4-01}
                    \begin{aligned}
                           W =& X_i f_i(Y_j) + W_1,\\
                        W_1 = & \mu_{i j k}(Y_l) X_i X_j A_k
                               + \nu_{i j k}(Y_l) X_i A_j A_k
                               + \xi_{i j}(Y_k) A_i A_j
                               + \kappa_{i j k}(Y_l) A_i A_j A_k +\\
                              &+ (\text{terms with more than three } X_i \text{'s and } A_i \text{'s}).
                    \end{aligned}
                \end{equation}
                Note again that each term of $W_1$ contains at least two $X_i$'s or $A_i$'s.  All the previous proof can be carried on to reach the revised Nelson-Seiberg theorem by considering the following two types of vacua:
                \begin{itemize}
                \item The discussion on vacua with $X_i = A_i = 0$ in the previous proof proceeds without change.  Any nonzero radius of convergence of the polynomial expansion~\eqref{eq:4-01} ensures the validity of such a vacuum at the origin.
                \item The discussion on vacua with $X_i \ne 0$ or $A_i \ne 0$ in the previous proof only involves the original Nelson-Seiberg theorem, which does not rely on the expansion form~\eqref{eq:4-01}.
                \end{itemize}
          \item If $W$ is singular at the origin, the vacuum, if existing, must be away from the origin to ensure a reliable effective theory calculation.  The R-symmetry is broken by some field expectation values at the vacuum, and SUSY is broken according to the original Nelson-Seiberg theorem.
          \end{itemize}
\end{itemize}

By identifying whether $W$ has a singularity at the origin of the field space, all cases with polynomial and nonpolynomial $W$'s are covered in our discussion.  Note that singularities away from the origin, if existing, do not affect the proofing process.  In summary, we have proved a generalized condition for SUSY breaking in models with generic superpotentials:

\begin{theorem}
(The Nelson-Seiberg theorem revised and generalized)
In a Wess-Zumino model with a generic superpotential, assuming the existence of a vacuum at the global minimum of the scalar potential, SUSY is spontaneously broken at the vacuum if and only if the superpotential has an R-symmetry, and one of the following conditions is satisfied:
\begin{itemize}
\item The superpotential is smooth at the origin of the field space, and the number of R-charge $2$ fields is greater than the number of R-charge $0$ fields for any possible consistent R-charge assignment.
\item The superpotential is singular at the origin of the field space.
\end{itemize}
\end{theorem}

Nonpolynomial superpotentials often appear as low energy effective descriptions of dynamical SUSY breaking models, which come from nonperturbative effects in supersymmetric quantum chromodynamics (SQCD) for various number of colors $N_c$ and number of flavors $N_f$~\cite{Intriligator:1995au, Shifman:1995ua, Shadmi:1999jy, Terning:2003th, Strassler:2003qg}.  A nonpolynomial Affleck-Dine-Seiberg (ADS) superpotential~\cite{Davis:1983mz, Affleck:1983mk} is generated from gaugino condensation in the case of $N_f < N_c - 1$, or from instantons in the case of $N_f = N_c - 1$.  As an example, the 3-2 model~\cite{Affleck:1984xz} has a gauge group $SU(3) \times SU(2)$, a global $U(1)$ symmetry and an R-symmetry $U(1)_R$, with the following chiral superfields:
\begin{equation} \label{eq:4-02}
Q:\ (3, 2)_{\frac{1}{3}, 1}, \quad
L:\ (1, 2)_{- 1, - 3}, \quad
\tilde{u}:\ (\bar{3}, 1)_{- \frac{4}{3}, - 8}, \quad
\tilde{d}:\ (\bar{3}, 1)_{\frac{2}{3}, 4},
\end{equation}
where the representations of $SU(3)$ and $SU(2)$ are written in parentheses, and the subscripts indicate $U(1)$ and $U(1)_R$ charges.  Assuming $SU(3)$ interactions are much stronger than $SU(2)$ interactions, the superpotential is
\begin{equation} \label{eq:4-03}
W = \frac{\Lambda_3^7}{Q Q \tilde{u} \tilde{d}} + \lambda Q \tilde{d} L.
\end{equation}
The first term is the ADS superpotential coming from SU(3) instantons, and the latter is the only renormalizable tree-level polynomial term respecting all symmetries.  If a vacuum does exist, the singular nonpolynomial term pushes field expectation values away from the origin, breaks the R-symmetry and SUSY according to the generalized theorem.  On the other hand, as a result of vanishing $SU(3)$ and $SU(2)$ D-terms, we can assume all fields have their vacuum expectation values of the same order $v$.  With an approximately minimal K\"ahler potential at weak coupling limit, the vacuum is calculated by minimizing the scalar potential $V = \lVert \partial_i W \rVert^2$.  Omitting constant coefficients, the vacuum expectation values of fields and $V$ are estimated to be
\begin{equation}
\langle Q \rangle \sim \langle L \rangle
                  \sim \langle \tilde{u} \rangle
                  \sim \langle \tilde{d} \rangle
                  \sim v
                  \sim \frac{\Lambda_3}{\lambda^{1/7}}, \quad
\langle V \rangle \sim \lambda^{10/7} \Lambda_3.
\end{equation}
Note that for simplicity, we used the same set of symbols for superfields~\eqref{eq:4-02} and their scalar components.  The nonzero $\langle V \rangle$ indicates a SUSY breaking vacua, which verifies the prediction of the theorem.

\section{Conclusion}

The generalized theorem proved in this work provides a tool to build SUSY models with R-symmetries and give either SUSY breaking or SUSY vacua.  For models smooth at the origin of the field space, one can arrange R-charges of fields to satisfy either $N_X > N_Y$ or $N_X \le N_Y$ and get the needed vacua.  For models singular at the origin of the field space, field expectation values are pushed away from the origin if a vacuum does exist, and SUSY is broken with a generic superpotential.  The theorem applies to both perturbative and nonperturbative models.  It allows one to efficiently survey a large number of different models without solving the F-term equations, and select the models with desired vacua to continue explicit model building.  It provides a guiding rule for low energy effective SUSY model building to study phenomenology of new physics beyond SM as well as string phenomenology.

The existence of a global minimum must be assumed in the proof of both the Nelson-Seiberg theorem and its revisions discussed in this work.  Indeed, there are models which have no minimum, only maxima, saddle points, and runaway directions asymptotically approaching SUSY at infinity~\cite{Ferretti:2007ec, Ferretti:2007rq, Azeyanagi:2012pc, Sun:2018hnk}.  The nonperturbative superpotential~\eqref{eq:4-03} of the 3-2 model, if lacking the requirement of vanishing D-terms, also has a runaway behavior.  These runaway directions may be lifted up by other effects such as D-terms, and give SUSY breaking vacua, just like what happens to the 3-2 model~\eqref{eq:4-03}.  Such lift-up mechanism can also be built into runaway models with polynomial superpotentials~\cite{Azeyanagi:2012pc}.  For phenomenology model building purpose, it is reasonable to classify such runaway models as the SUSY breaking type, which can also be covered by both the Nelson-Seiberg theorem and its revisions discussed in this work.

It should be noted that nonperturbative effects do not necessarily lead to nonpolynomial superpotentials.  The form of the superpotential depends on how it is parameterized.  For example, in the Kachru-Kallosh-Linde-Trivedi (KKLT) construction for de Sitter vacua in type IIB flux compactifications~\cite{Kachru:2003aw}, The superpotential
\begin{equation}
W = W_0 + W_{\text{corr}} = W_0 + A e^{i a \rho}
\end{equation}
has a tree level contribution $W_0$ from fluxes, and a nonperturbative correction $W_{\text{corr}}$ from D3 brane instantons or gaugino condensation of the gauge theory on a stack of D7 branes, which stabilizes the volume modulus $\rho$.  The exponential form of $W_{\text{corr}}$ is smooth at any value of $\rho$, and no R-charge can be consistently assigned to $\rho$.  An supersymmetric anti-de Sitter vacuum is found at finite $\rho$ by minimizing the supergravity scalar potential with a nonminimal K\"ahler potential.  This model is base on supergravity, thus lies out of the scope of all theorems discussed in this work.

\section*{Acknowledgement}

We thank Yan He for helpful discussions.  This work is supported by the National Natural Science Foundation of China under grant 11305110.

\end{document}